\documentclass[usenatbib,usegraphicx]{mn2e}
\def\msun{{\rm M}_{\odot}}
\def\rsun{{\rm R}_{\odot}}

\def\micron{{\rm \mu m}}

\def\spitzer{{\it Spitzer\ }}

\begin{document}

\title[Rotation of low-mass stars in NGC 2547]{The Monitor project:
    Rotation of low-mass stars in the open cluster NGC 2547}
\author[J.~M.~Irwin et al.]{Jonathan~Irwin$^{1,2}$\thanks{E-mail: jmi at
    ast.cam.ac.uk}, Simon~Hodgkin$^{1}$, Suzanne~Aigrain$^{3}$,
    Jerome~Bouvier$^{4}$,
\newauthor
Leslie~Hebb$^{5}$ and Estelle~Moraux$^{4}$ \\
$^{1}$Institute of Astronomy, University of Cambridge, Madingley Road,
  Cambridge, CB3 0HA, United Kingdom \\
$^{2}$Harvard-Smithsonian Center for Astrophysics, 60 Garden Street,
    Cambridge, MA 02138, USA \\
$^{3}$Astrophysics Group, School of Physics, University of Exeter, Stocker Road,
  Exeter, EX4 4QL, United Kingdom \\
$^{4}$Laboratoire d'Astrophysique, Observatoire de Grenoble, BP 53,
  F-38041 Grenoble C\'{e}dex 9, France \\
$^{5}$School of Physics and Astronomy, University of St Andrews,
  North Haugh, St Andrews, KY16 9SS, Scotland}
\date{}

\maketitle

\begin{abstract}
We report on the results of an $I$-band time-series photometric
survey of NGC 2547 using the MPG/ESO 2.2m telescope with WFI,
achieving better than $1\%$ photometric precision per data point over
$14 \la I \la 18$.  Candidate cluster members were selected
from a $V$ vs $V-I$ colour magnitude diagram over $12.5 < V < 24$
(covering masses from $0.9\ \msun$ down to below the brown dwarf
limit), finding $800$ candidates, of which we expect $\sim 330$ to
be real cluster members, taking into account contamination from the
field (which is most severe at the extremes of our mass range).
Searching for periodic variations in these gave $176$ detections over
the mass range $0.1 \la M/\msun \la 0.9$.  The rotation period
distributions were found to show a clear mass-dependent morphology,
qualitatively intermediate between the distributions obtained from
similar surveys in NGC 2362 and NGC 2516, as would be expected from
the age of this cluster.  Models of the rotational evolution were
investigated, finding that the evolution from NGC 2362 to NGC 2547 was
qualitatively reproduced (given the uncertainty in the age of NGC
2547) by solid body and core-envelope decoupled models from our
earlier NGC 2516 study without need for significant modification.
\end{abstract}

\section{Introduction}
\label{intro_section}

NGC 2547 is a relatively well-studied nearby young open cluster; we
adopt the parameters of \citet{nj2006}, who give an age of
$38.5^{+3.5}_{-6.5}\ {\rm Myr}$, distance modulus $(M - m)_0 =
7.79^{+0.11}_{-0.05}$ and reddening $A_V = 0.186$.  The age derived by
these authors puts the cluster at the approximate point where G-dwarfs
reach the zero age main sequence (ZAMS; see for example Figure 1 of
\citealt{i2007b}), and slightly younger than the typically-used
benchmark of $\alpha$ Persei ($\sim 50\ {\rm Myr}$,
e.g. \citealt{bm99}).  The small distance modulus makes it possible to
probe down to very low masses, with relatively modest-sized telescopes.

The cluster has been surveyed in X-rays using ROSAT \citep{jt98} and
XMM-Newton \citep{j2006}.  The latter study found that the solar type
stars in NGC 2547 exhibited a similar relation between X-ray activity
and Rossby number to field stars and older clusters, but with
saturated or super-saturated X-ray activity levels, and median levels
of X-ray luminosity and X-ray to bolometric luminosity ratio similar
to T-Tauri stars in the ONC, but much higher than in the Pleiades.
These measures are consistent with the relative youth of NGC 2547.

Surveys of lithium depletion (\citealt{o2003}; \citealt{j2003};
\citealt{j2005}) have placed constraints on the cluster age, and
conversely, on stellar evolution models by comparison with isochrone
fitting results.  \citet{j2005} find ages based on lithium depletion
of $34 - 36\ {\rm Myr}$, reasonably consistent with the
isochrone-based estimate we assumed earlier.

\citet{jtj2000} performed a $v \sin i$ survey of the solar-type stars
in the cluster, finding a distribution indistinguishable from that in
the similar-age clusters IC 2391 and IC 2602, with a similar
rotation-activity relationship to older clusters such as the Pleiades.
\citet{l2003} found weak evidence for mass segregation in the cluster,
confirmed by \citet{j2003} using a new photometric survey, where they
also derived the cluster mass function down to $\sim 0.05\ \msun$,
finding a result very similar to the Pleiades mass function over the
range $0.075 < M/\msun < 0.7$.

Finally, NGC 2547 has been observed with \spitzer using the IRAC
and MIPS instruments by \citet{y2004} to investigate the cluster disc
frequency.  \citet{gorlova07} subsequently extended the survey
coverage, finding only $2-4$ out of $\sim 600$ stars of B---mid-M
spectral type to show excesses in the IRAC $8\ \micron$ band, a
fraction of $\la 1$ per cent.  In the MIPS $24\ \micron$ band,
they found $\sim 40$ per cent of the stars of B---F spectral types to
show excesses.  This striking result is most likely due to the discs
clearing from the inside outwards, with the lack of short-wavelength
excesses caused by the discs being substantially cleared within the
central $\sim 1\ {\rm AU}$ parts by the age of NGC 2547.

\subsection{Evolution of stellar angular momentum}
\label{amevol_section}

For a discussion of the context of this work, the reader is referred
to our M34 and NGC 2516 publications in \citet{i2006} and
\citet{i2007b}.  The present survey probes a different open cluster of
younger age ($\sim 40\ {\rm Myr}$ for NGC 2547, compared to $\sim 150\
{\rm Myr}$ for NGC 2516 and $200\ {\rm Myr}$ for M34).  Due to the
small distance modulus and young age, the NGC 2547 survey probes a
relatively wide range of masses, covering $0.1 \la M/\msun \la 0.9$,
compared to $0.1 \la M/\msun \la 0.7$ in NGC 2516 and $0.25 \la
M/\msun \la 1.0$ in M34.  This will allow a more detailed comparison
with models of angular momentum evolution, and in particular to test
our results in \citet{i2007b}, where we found that the evolution of
the fast rotators appeared to be better-reproduced by solid body
models, and the slow rotators by differentially rotating models.  At
the age of NGC 2547, stars with masses $M \la 0.4\ \msun$ should be
fully-convective, and at $0.4\ \msun$, just beginning to develop a
radiative core \citep{cb97}.

\subsection{The survey}

We have undertaken a photometric survey in NGC 2547 using the MPG/ESO
2.2m telescope with the Wide Field Imager (WFI; \citealt{b99}).  Our
goals are two-fold: first, to study rotation periods for a sample of
low-mass members, covering K and M spectral types, down to $\sim 0.1\
\msun$, and second, to look for eclipsing binary systems containing
low-mass stars, to obtain dynamical mass measurements in conjunction
with radial velocities from follow-up spectroscopy.  Such systems
provide the most accurate determinations of fundamental stellar
parameters (in particular, masses) for input to models of stellar
evolution, which are poorly constrained in this age range.  We defer
discussion of our eclipsing binary candidates to a later paper once we
have obtained suitable follow-up spectroscopy.

These observations are part of a larger photometric monitoring
survey of young open clusters over a range of ages and metalicities
(the Monitor project; \citealt{hodg06} and \citealt{a2007}).

The remainder of the paper is structured as follows: the observations
and data reduction are described in \S \ref{odr_section}, and the
colour magnitude diagram (CMD) of the cluster and candidate membership
selection are presented in \S \ref{memb_section}.  The method we use
for obtaining photometric periods is summarised in \S
\ref{period_section} (see \citealt{i2006} for a more detailed
discussion), and our results are summarised in \S
\ref{results_section}.  We discuss the implications of these results in
\S \ref{disc_section}, and our conclusions are summarised in \S
\ref{conclusions_section}.

\section{Observations and data reduction}
\label{odr_section}

Photometric monitoring observations were obtained using the MPG/ESO
2.2m telescope with WFI in service mode, with $\sim 100\ {\rm hours}$
of observations distributed over a $\sim 8\ {\rm month}$ period from
2005 October to 2006 May.  The instrument provides a field of view of
$\sim 34' \times 33'$ ($0.31\ {\rm sq. deg}$), using a mosaic of eight
${\rm 2k} \times {\rm 4k}$ pixel CCDs, at a scale of $\sim 0.238'' /
{\rm pix}$.

\begin{figure}
\centering
\includegraphics[angle=0,width=3.2in]{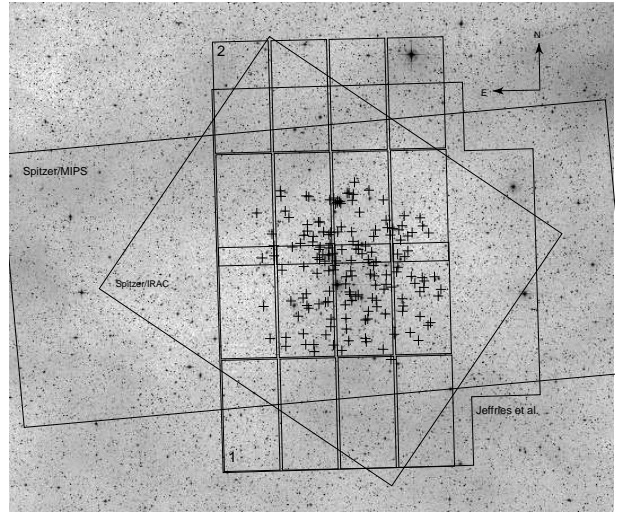}

\caption{Digitised sky survey (DSS) image of NGC 2547 covering $\sim
  1.4^\circ \times 1.3^\circ$, showing the coverage of the present
  survey (numbered 8-chip mosaic tiles), the \spitzer IRAC and MIPS
  fields of \citet{y2004}, and the optical survey of \citet{j2004}.
  Crosses show the positions of the X-ray sources from \citet{j2006}.}

\label{coverage}
\end{figure}

\begin{figure}
\centering
\includegraphics[angle=270,width=3.2in]{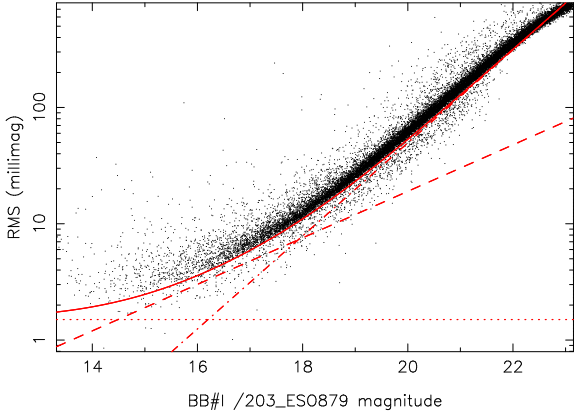}

\caption{Plot of RMS scatter as a function of magnitude for the
$I$-band observations of a single field in NGC 2547, for all unblended
objects with stellar morphological classifications.  The diagonal
dashed line shows the expected RMS from Poisson noise in the object,
the diagonal dot-dashed line shows the RMS from sky noise in the
photometric aperture, and the dotted line shows an additional $1.5\
{\rm mmag}$ contribution added in quadrature to account for systematic
effects.  The solid line shows the overall predicted RMS, combining
these contributions.}

\label{rmsplot}
\end{figure}

In order to maximise the number of cluster members covered by our
survey we elected to use two fields, with a $120\ {\rm s}$ $I$-band
exposure in each, observed by cycling between them, to give
a cadence of $\sim 7\ {\rm minutes}$, covering $\sim 0.6\ {\rm
  sq. deg}$ of the cluster, illustrated in Figure \ref{coverage}.  Our
observations are sufficient to give $1 \%$ or better photometric
precision per data point from saturation at $I \sim 14$ down to $I
\sim 18$ (see Figure \ref{rmsplot}), covering K to mid M spectral
types at the age and distance of NGC 2547.

We also obtained $120\ {\rm s}$ and $1500\ {\rm s}$ $V$-band exposures
in each field during photometric conditions, to generate a colour
magnitude diagram (CMD).

For a full description of our data reduction steps, the reader is
referred to \citet{i2007a}.  Briefly, we used the pipeline for
the INT wide-field survey \citep{il2001} for 2-D 
instrumental signature removal (bias correction, flatfielding,
defringing) and astrometric and photometric calibration.  We then 
generated a master catalogue for each filter by stacking $20$
of the frames taken in the best conditions (seeing, sky brightness and
transparency) and running the source detection software on the stacked
image.  The resulting source positions were used to perform aperture
photometry on all of the time-series images.  We achieved a per data
point photometric precision of $\sim 2-4\ {\rm mmag}$ for the
brightest objects, with RMS scatter $< 1 \%$ for $I \la 18$ (see
Figure \ref{rmsplot}).

Our source detection software flags any objects detected as having
overlapping isophotes.  This information is used, in conjunction with
a morphological image classification flag also generated by the
pipeline software \citep{il2001} to allow us to identify non-stellar
or blended objects in the time-series photometry.

Photometric calibration of our data was carried out using regular
observations of \citet{l92} equatorial standard star fields in the
usual way, as part of the standard ESO nightly calibrations.

Light curves were extracted from the data for $\sim 130\,000$ objects,
$90\,000$ of which had stellar morphological classifications, using our
standard aperture photometry techniques, described in \citet{i2007a}.
We fit a 2-D quadratic polynomial to the residuals in each frame 
(measured for each object as the difference between its magnitude on
the frame in question and the median calculated across all frames) as
a function of position, for each of the $8$ CCDs separately.
Subsequent removal of this function accounted for effects such as
varying differential atmospheric extinction across each frame.  Over a
single CCD, the spatially-varying part of the correction remains
small, typically $\sim 0.02\ {\rm mag}$ peak-to-peak.  The reasons for
using this technique are discussed in more detail in \citet{i2007a}.

For the production of deep CMDs, we used the $I$-band master
catalogue, and combined the $V$-band exposures.  The limiting
magnitudes, measured as the approximate magnitude at which our
catalogues are $50\%$ complete on these images were $V \sim 23$
and $I \sim 23$.

\section{Selection of candidate low-mass members}
\label{memb_section}

A catalogue of photometrically-selected candidate members (using $R I
Z$ photometry) covering most of our survey area was available from
\citet{j2004} for solar mass down to $\sim 0.05\ \msun$ (limiting
magnitude $I \simeq 19.5$), but we elected to perform a new
photometric selection using $V$ versus $V-I$ CMDs from our data, to
make better use of the deeper limiting magnitudes of the present
survey.

\subsection{The $V$ versus $V - I$ CMD}
\label{cmd_section}

\begin{figure}
\centering
\includegraphics[angle=270,width=3.5in]{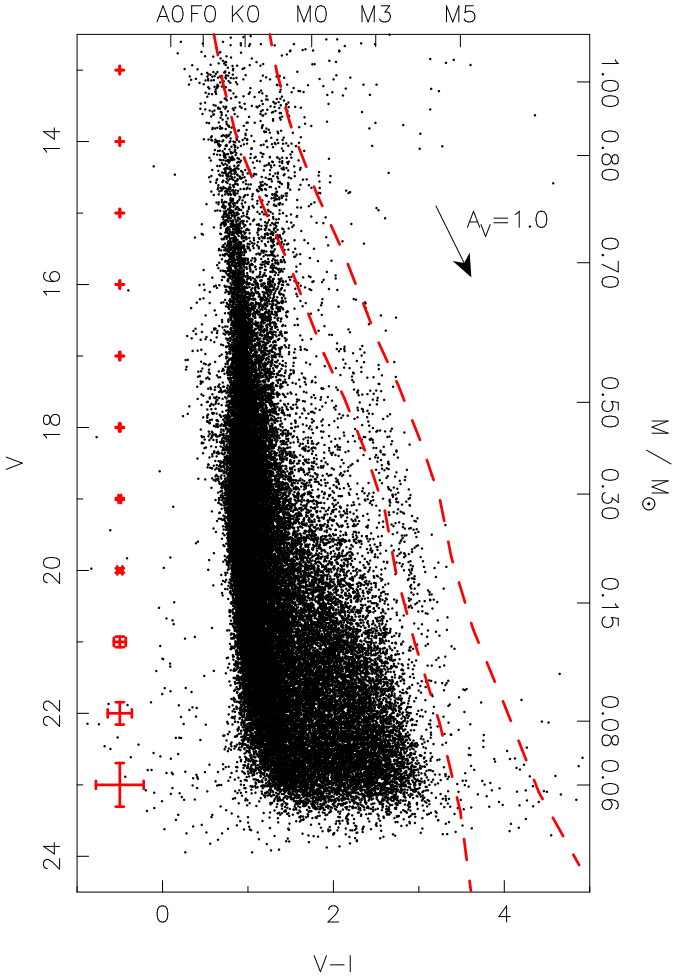}

\caption{$V$ versus $V - I$ CMD of NGC 2547 from stacked images, for
all objects with stellar morphological classification.  The cluster
sequence is clearly visible on the right-hand side of the diagram.
The boundaries of the region used to select photometric candidate
members are shown by the dashed lines (all objects between the dashed
lines were selected).  The reddening vector for $A_V = 1.0$ is shown
at the right-hand side of the diagram.  The mass scale is from the
$40\ {\rm Myr}$ NextGen models \citep{bcah98} for $M > 0.1\ \msun$,
and the $50\ {\rm Myr}$ DUSTY models \citep{cbah2000} for $M < 0.1\
\msun$, using our empirical isochrone to convert the $V$ magnitudes to
$I$ magnitudes, and subsequently obtaining the masses from these, due
to known problems with the $V$ magnitudes from the models (see \S
\ref{cmd_section}).  The error bars at the left-hand side of the plot
indicate the typical photometric error for an object on the cluster
sequence.}

\label{cmd}
\end{figure}

Our CMD of NGC 2547 is shown in Figure \ref{cmd}.  The $V$ and $I$
measurements were converted to the standard Johnson-Cousins
photometric system using colour equations derived from standard
star observations by Mike Irwin (private communication):
\begin{eqnarray}
(V - I)& = &(V_{ccd} - I_{ccd})\ /\ 1.14 \\
V& = &V_{ccd} - 0.08\ (V - I) \\
I& = &I_{ccd} + 0.06\ (V - I)
\end{eqnarray}

Candidate cluster members were selected by defining an empirical
cluster sequence `by eye' to follow the clearly-visible cluster
single-star sequence.  The cuts were defined by moving this line along
a vector perpendicular to the cluster sequence, by amounts $k -
\sigma(V - I)$ and $k + \sigma(V - I)$ as measured along this vector,
where $\sigma(V - I)$ is the photometric error in the $V - I$ colour.
The values of $k$ used were $-0.15\ {\rm mag}$ for the lower line and
$0.5\ {\rm mag}$ for the upper line on the diagram, making the
brighter region wider to avoid rejecting binary and multiple systems,
which are overluminous for their colour compared to single stars.
$800$ candidate photometric members were selected, over the full $V$
magnitude range from $V = 12.5$ to $24$, but the well-defined cluster
sequence appears to terminate at $M \sim 0.1\ \msun$, or $V \sim 21$,
with a few candidate members below this limit, but with high field
contamination.

The selection region used here is comparable to that of \citet{j2004},
and below $V \sim 16$, $100\%$ of their candidate cluster members are
also selected by our cut.  Above $V \sim 16$, there are slight
differences in the isochrones and due to saturation in the present
sample, and the corresponding fraction is reduced to $\sim 90\%$.

We also considered using the model isochrones of \citet{bcah98} and
\citet{cbah2000} for selecting candidate members.  The NextGen model
isochrones were found to be unsuitable due to the known discrepancy
between these models and observations in the $V - I$ colour for
$T_{\rm eff} \la 3700\ {\rm K}$ (corresponding here to $V - I \ga 2$).
This was examined in more detail by \citet{bcah98}, and is due to a
missing source of opacity at these temperatures, leading to
overestimation of the $V$-band flux.  Consequently, when we have used
the NextGen isochrones to determine model masses and radii for our
objects, the $I$-band absolute magnitudes were used to perform the
relevant look-up, since these are less susceptible to the missing
source of opacity, and hence give more robust estimates.

\subsection{Contamination}
\label{contam_section}

\begin{figure}
\centering
\includegraphics[angle=270,width=3in]{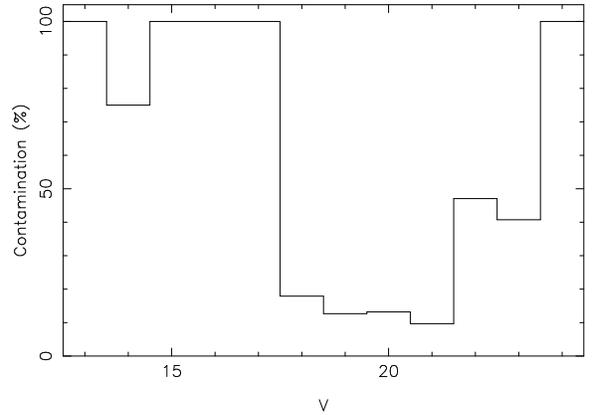}

\caption{Contamination, measured as the ratio of the calculated number
  of objects in each magnitude bin from the Galactic models, to the
  number of objects detected and classified as candidate cluster
  members in that magnitude bin.  Note that bins with contamination
  estimates $> 100\%$ (where there were more objects in that bin from
  the Galactic model than were actually observed) have been truncated
  to $100\%$.}

\label{contam}
\end{figure}

In order to estimate the level of contamination in our catalogue, we
used the Besan\c{c}on Galactic models \citep{r2003} to generate a
simulated catalogue of objects passing our selection criteria at the
Galactic coordinates of NGC 2547 ($l = 264.4^\circ$, $b = -8.5^\circ$),
covering the total FoV of $\sim 0.59\ {\rm sq.deg}$ (including gaps
between detectors).  We selected all objects over the apparent
magnitude range $10 < V < 24$, giving $120\,000$ stars.  The same
selection process as above for the cluster members was then applied to
find the contaminant objects.  A total of $888$ simulated objects
passed these membership selection criteria, giving an overall
contamination level of $\sim 59 \%$ after correcting for bins where
the number of objects predicted by the models exceeded the number
actually observed (we simply assumed $100 \%$ field contamination in
these bins).  Figure \ref{contam} shows the contamination as a
function of $V$ magnitude.  We note that this figure is somewhat
uncertain due to the need to use Galactic models, and especially given
the overestimation of the numbers of observed objects by the models.
Follow-up data will be required to make a more accurate estimate.

\section{Period detection}
\label{period_section}

\subsection{Method}
\label{method_section}

The method we use for detection of periodic variables is described in
detail in \citet{i2006}, and we provide only a brief
summary here.  The method uses least-squares fitting of sine curves to
the time series $m(t)$ (in magnitudes) for {\em all} candidate cluster
members, using the form:
\begin{equation}
m(t) = m_{dc} + \alpha \sin(\omega t + \phi)
\label{sine_eqn}
\end{equation}
where $m_{dc}$ (the DC light curve level), $\alpha$ (the amplitude) and
$\phi$ (the phase) are free parameters at each value of $\omega$ over
an equally-spaced grid of frequencies, corresponding to periods from
$0.005 - 100\ {\rm days}$ for the present data-set.

Periodic variable light curves were selected by evaluating the change
in reduced $\chi^2$:
\begin{equation}
\Delta \chi^2_\nu = \chi^2_\nu - \chi^2_{\nu,{\rm smooth}} > 0.4
\end{equation}
where $\chi^2_\nu$ is the reduced $\chi^2$ of the original light curve
with respect to a constant model, and $\chi^2_{\nu,{\rm smooth}}$ is the
reduced $\chi^2$ of the light curve with the smoothed, phase-folded
version subtracted.  This threshold was used for the M34 data and
appears to work well here too, carefully checked by examining all the
light curves for two of the detectors, chosen randomly.  A total of
$482$ objects were selected by this automated part of the procedure.

The selected light curves were examined by eye, to define the final
sample of periodic variables.  A total of $176$ light curves were
selected, with the remainder appearing non-variable or too ambiguous
to be included.

\subsection{Simulations}
\label{sim_section}

Monte Carlo simulations were performed following the method detailed
in \citet{i2006}, injecting simulated signals of $2 \%$ amplitude and
periods chosen following a uniform distribution on $\log_{10}$ period
from $0.1$ to $20\ {\rm days}$, into light curves covering a uniform
distribution in mass, from $1.0$ to $0.1\ \msun$.  A total of $1015$
objects were simulated.

\begin{figure*}
\centering
\includegraphics[angle=270,width=6in]{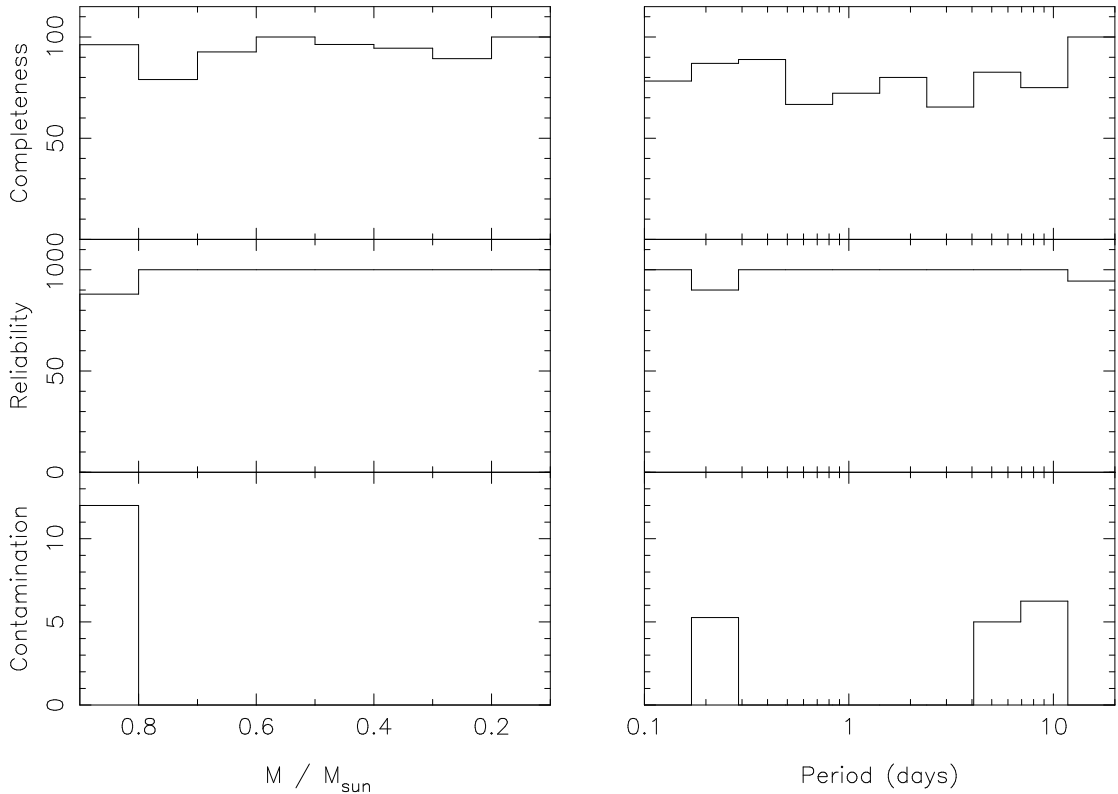}

\caption{Results of the simulations for $0.02\ {\rm mag}$ amplitude
  expressed as percentages, plotted as a function of mass (left) and
  period (right).  The simulated region covered $0.1 < {\rm M}/\msun <
  0.9$ in order to be consistent with the NGC 2547 sample.  {\bf Top
  panels}: completeness as a function of real (input) period.  {\bf
  Centre panels}: Reliability of period determination, plotted as the
  fraction of objects with a given true period, detected with the
  correct period (defined as differing by $< 20\%$ from the true
  period).  {\bf Bottom panels}: Contamination, plotted as the
  fraction of objects with a given detected period, having a true
  period differing by $> 20\%$ from the detected value.}

\label{sim_results}
\end{figure*}

\begin{figure}
\centering
\includegraphics[angle=270,width=3in,bb=59 107 581 630,clip]{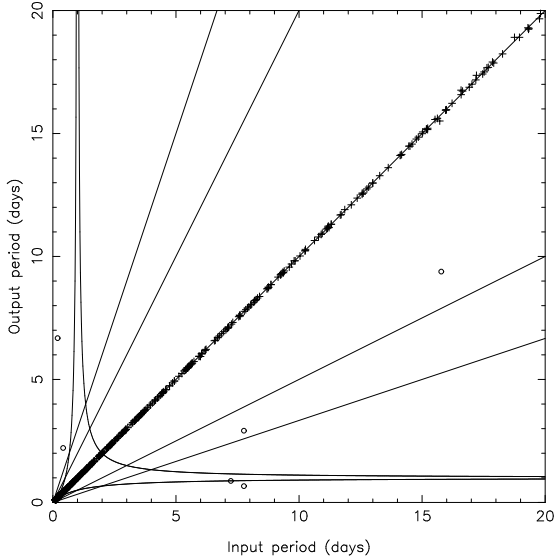}

\caption{Detected period as a function of actual (input) period for our
  simulations.  Objects plotted with crosses had fractional period
  error $< 10\%$, open circles $> 10\%$.  The straight lines represent
  equal input and output periods, and factors of $2$, $3$, $1/2$ and
  $1/3$.  The curved lines are the loci of the $\pm 1\ {\rm day^{-1}}$
  aliases resulting from gaps during the day.  The majority of the
  points fall on (or close to) the line of equal periods.}

\label{periodcomp}
\end{figure}

The results of the simulations are shown in Figure \ref{sim_results}
as diagrams of completeness, reliability and contamination
as a function of period and stellar mass.  Broadly, our period
detections are close to $100 \%$ complete from $0.9\ \msun$ down to
$0.1\ \msun$, with remarkably little period dependence.
Figure \ref{periodcomp} shows a comparison of the detected periods
with real periods for our simulated objects, indicating remarkably
high reliability, especially compared to the earlier data-sets 
(\citealt{i2006}; \citealt{i2007b}), resulting from the extremely long
time base-line of the present survey afforded by observing in service
mode.

These results indicate overall that the service mode observing
strategy as implemented by ESO performs very favourably for rotation
period detection in NGC 2547, compared to our conventional visitor
mode observations (e.g. \citealt{i2006}; \citealt{i2007b}).  In
particular, the sampling is very well-suited to determining extremely
precise and accurate rotation periods, as illustrated in Figure
\ref{periodcomp}.

However, there are two principal disadvantages.  The long time span of
the observations, and sparse sampling, here over $\sim 8\ {\rm
  months}$, means the process of period detection is complicated by
the evolution of the spot patterns on the stellar surfaces giving rise
to the photometric modulations.  This has the effect of modifying the
amplitude and phase of the variability, and may occur over timescales
of a few months.  This has not been accounted for in the simulations.
Qualitatively, this effect may reduce the overall detection rate since
the variations will not produce a smooth phase-folded light curve for
any plausible period, and will therefore be rejected in the detection
process.  Furthermore, for eclipsing binary detection this is obviously
a serious problem, since the variations become much more difficult to
remove when searching for shallow eclipses.

The second problem relates to the observing conditions.  Since desired
weather conditions are specified for a service mode proposal, and
differential photometry does not require the best observing
conditions, a service mode programme will in general receive poorer
conditions than a visitor mode programme performed in good weather
(e.g. the NGC 2516 survey; \citealt{i2007b}).  For rotation period
detection at reasonable amplitudes, the simulations indicate that this
is not a serious issue.  However, poor seeing and particularly poor
transparency, increase the level of low-amplitude correlated noise in
the light curves (see \citealt{i2007a}), which severely impedes
detection of shallow transit events.

The effects of both these factors on transit detection will be
evaluated in more detail in a future publication presenting our
eclipse candidates.

\subsection{Detection rate and reliability}

\begin{figure}
\centering
\includegraphics[angle=270,width=3.5in]{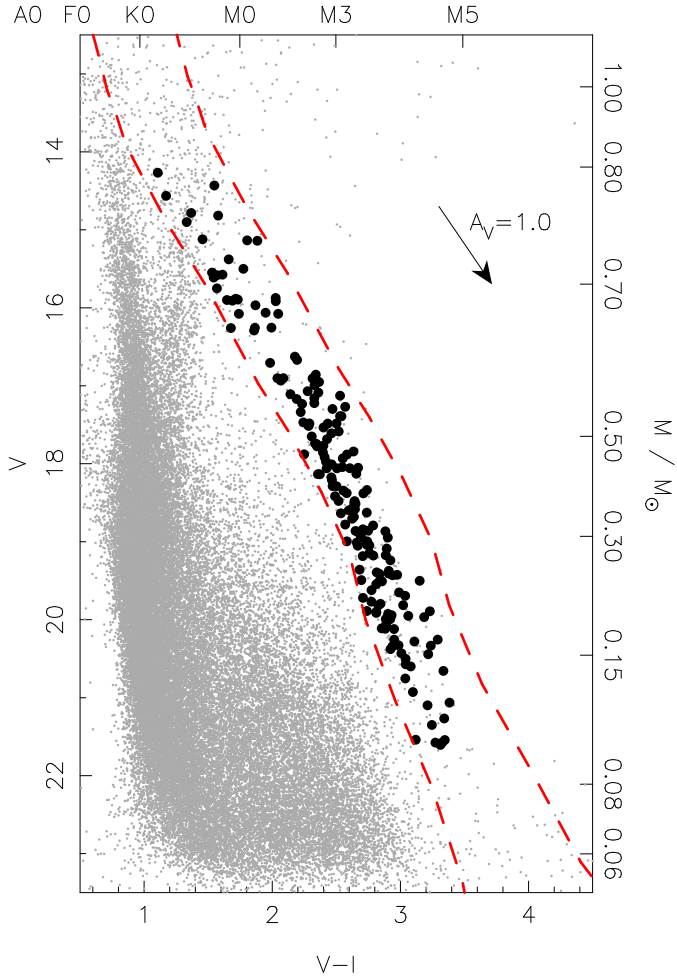}

\caption{Magnified $V$ versus $V - I$ CMD of NGC 2547, for objects
with stellar morphological classification, as Figure \ref{cmd},
showing all $176$ candidate cluster members with detected periods
(black points).  The dashed lines show the cuts used to select
candidate cluster members (see \S \ref{cmd_section}).}

\label{cands_on_cmd}
\end{figure}

The locations of our detected periodic variable candidate cluster
members on a $V$ versus $V-I$ CMD of NGC 2547 are shown in Figure
\ref{cands_on_cmd}.  The diagram indicates that the majority of
the detections lie on the single-star cluster sequence, as would
be expected for rotation in cluster stars as opposed to, say,
eclipsing binaries.

\begin{figure}
\centering
\includegraphics[angle=270,width=3in]{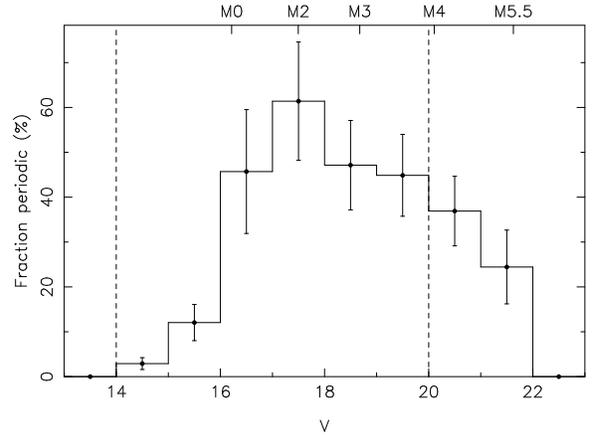}

\caption{Fraction of candidate cluster members detected as periodic
  variables, plotted as a function of magnitude.  This distribution
  has not been corrected for incompleteness in the period detections,
  which are close to $100\%$ complete for $14 < V < 20$ (shown by the
  vertical dashed lines), or for the effects of field contamination.}

\label{fracper}
\end{figure}

Figure \ref{fracper} shows the fraction of cluster members with
detected periods as a function of $V$ magnitude.  The decaying parts
of the histogram at the bright and faint ends may be caused by the
increased field contamination here (see Figure \ref{contam}), since we
expect field objects on average show less rotational modulation than
cluster objects, and/or incompleteness effects resulting from
saturation for $V \la 14$, and for $V \ga 20$, the gradual increase in
the minimum amplitude of variations we can detect (corresponding to
the reduction in sensitivity moving to fainter stars, see Figure
\ref{rmsplot}).

In the M34 survey of \citet{i2006}, we commented on a possible
increase in the fraction of photometric variables from $K$ to $M$
spectral types, subject to a large uncertainty due to small number
statistics.  This feature was also seen in the NGC 2516 survey
\citep{i2007b}, and the three distributions look qualitatively very
similar after correcting for the different mass ranges covered, with a
rise in the fraction of periodic objects detected at spectral types of
$\sim$ M0.  However, in all three cases, this is also the point at
which the cluster sequence becomes well-separated from the field in
the CMD analysis, corresponding to a reduction in field contamination
moving toward later spectral types.  It is therefore difficult to say
if the trend is significant from the present observations.

Likewise, lower field contamination is expected in the rotation sample
than in the full candidate membership sample.  Typical field
population ages for the young disc of $3\ {\rm Gyr}$ (\citealt{may74};
\citealt{msr91}) imply slower rotation rates by factors of $\sim$ a
few than cluster members, and reduced activity, implying smaller
asymmetric components of the spot coverage and hence lower photometric
amplitudes (which probably render many of them undetectable).

The properties of all our rotation candidates are listed in Table
\ref{cand_table}.

\begin{table*}
\centering
\begin{tabular}{lllrrrrrrrr}
\hline
Identifier     &RA    &Dec   &$V$   &$I$   &$P$    &$\alpha_i$ &$M$ &$R$ \\
               &J2000 &J2000 &mag   &mag   &days   &mag        &$\msun$ &$\rsun$ \\
\hline
N2547-1-1-2501 &08 11 20.69 &-48 49 22.5 &17.97 &15.55 & 2.758 &0.014 &0.44 &0.56 \\
N2547-1-1-3141 &08 11 32.39 &-48 48 01.0 &17.78 &15.43 & 9.523 &0.009 &0.47 &0.59 \\
N2547-1-1-3640 &08 11 59.66 &-48 46 56.8 &15.58 &14.01 & 3.213 &0.038 &0.76 &0.82 \\
N2547-1-1-3903 &08 11 22.55 &-48 46 29.8 &14.90 &13.57 & 0.991 &0.005 &0.81 &0.89 \\
N2547-1-1-4637 &08 11 46.91 &-48 44 54.0 &17.16 &14.83 & 3.071 &0.024 &0.61 &0.69 \\
\hline
\end{tabular}

\caption{Properties of our $176$ rotation candidates.
  The period $P$ in days, $i$-band amplitude $\alpha_i$ (units of
  magnitudes, in the instrumental bandpass), interpolated mass and
  radius (from the models of \citealt{bcah98}, derived using the $I$
  magnitudes) are given (where available).  Our identifiers are formed
  using a simple scheme of the cluster name, field number, CCD number
  and a running count of stars in each CCD, concatenated with dashes.
  The full table is available in the electronic edition.  Machine
  readable copies of the data tables from all the Monitor rotation
  period publications are also available at {\tt
  http://www.ast.cam.ac.uk/research/monitor/rotation/}.}
\label{cand_table}
\end{table*}

\subsection{Non-periodic objects}

The population of objects rejected by the period detection procedure
described in \S \ref{period_section} was examined, finding that the
most variable population of these light curves (which might correspond
to non-periodic or semi-periodic variability) was contaminated by a
small number of light curves ($\sim 50$) exhibiting various uncorrected
systematic effects, mostly seeing-correlated variations due to image
blending.  It is therefore difficult to quantify the level of
non-periodic or semi-periodic variability in NGC 2547 from our
data.  Qualitatively however, there appear to be very few of these
variables, and examining the light curves indicated only $\sim 10$
obvious cases, some of which resembled eclipses (presumably due to
eclipsing binaries), and will be the subject of a later Monitor
project paper.

\section{Results}
\label{results_section}

\subsection{NGC 2547 rotation periods}
\label{prv_section}

\begin{figure}
\centering
\includegraphics[angle=270,width=3in]{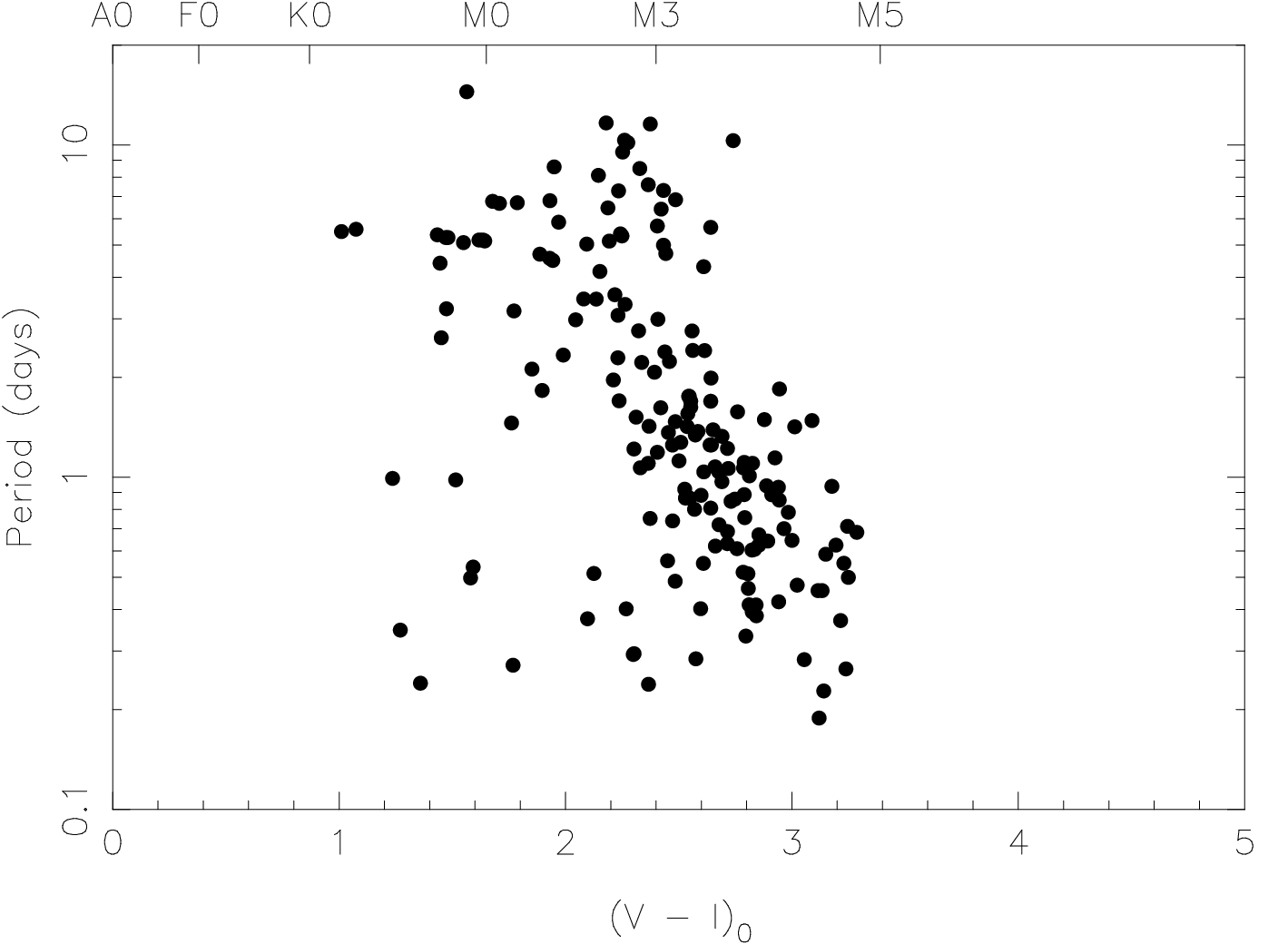}
\includegraphics[angle=270,width=3in]{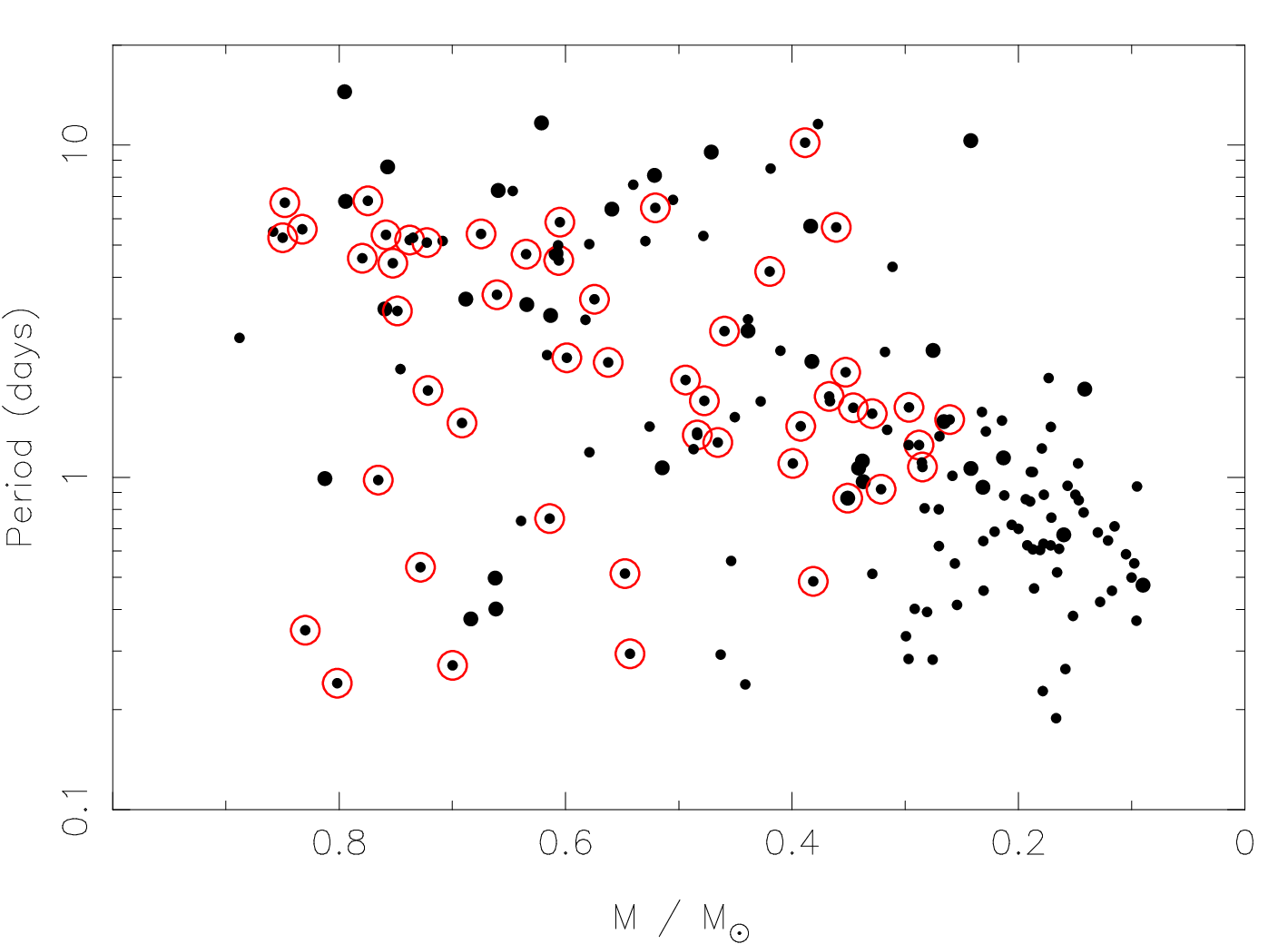}

\caption{Plots of rotation period as a function of dereddened $V-I$
  colour (top), and mass (bottom) for NGC 2547, deriving the masses
  using the $40\ {\rm Myr}$ NextGen mass-magnitude relations of
  \citet{bcah98} and the measured $I$-band magnitudes.}

\label{pcd}
\end{figure}

Plots of period as a function of $V-I$ colour and mass for the objects
photometrically selected as possible cluster members are shown in
Figure \ref{pcd}.  These diagrams show a striking correlation between
stellar mass (or spectral type) and the longest rotation period seen
at that mass, with a clear lack of slow rotators at very low masses.
This trend is also followed by the majority of the rotators, with only
a tail of faster rotators to $\sim 0.2\ {\rm day}$ periods.
Furthermore, very few objects were found rotating faster than this,
implying a hard lower limit to the observed rotation periods at $0.2\
{\rm days}$.

\begin{figure}
\centering
\includegraphics[angle=270,width=3in]{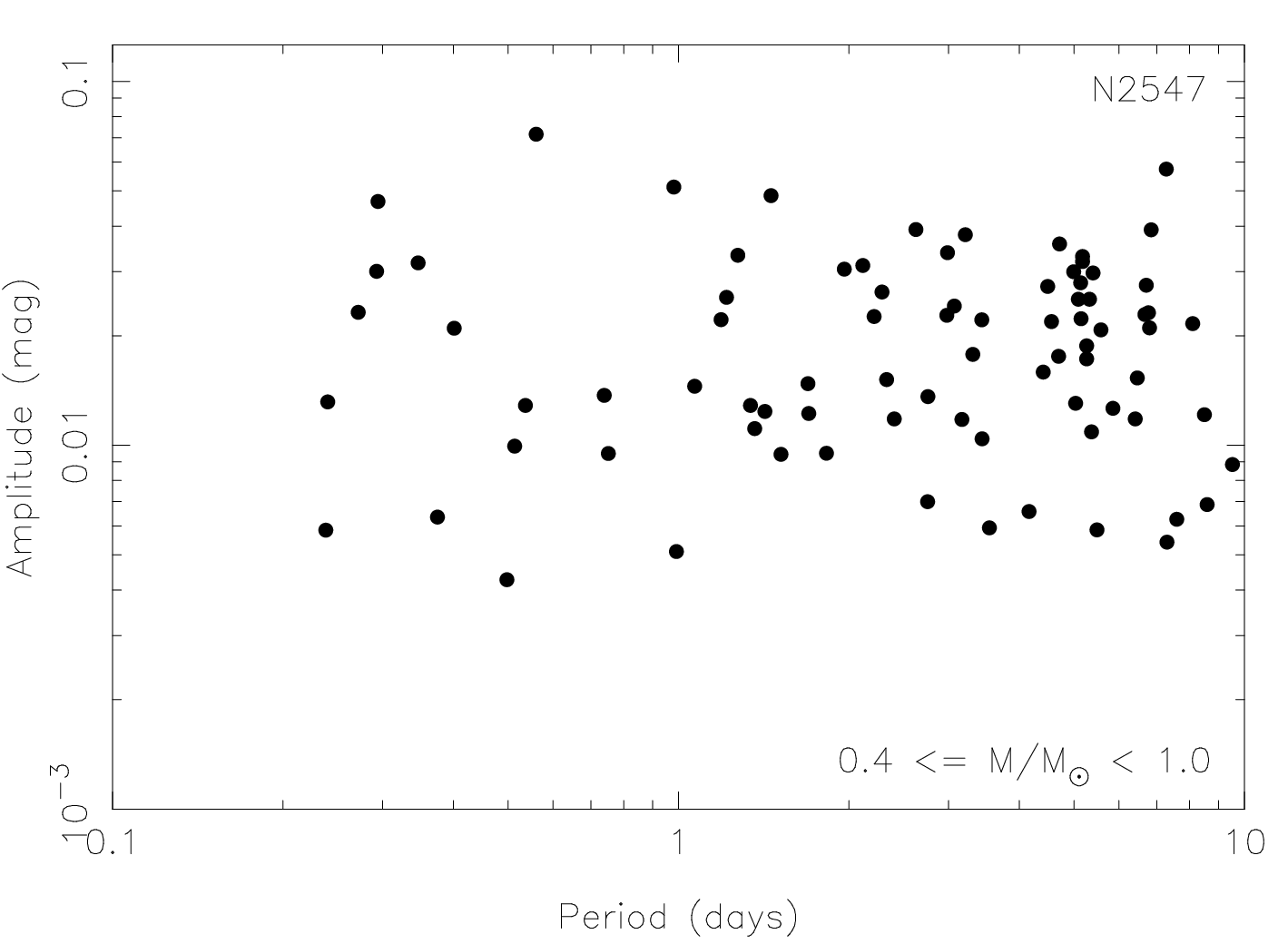}
\includegraphics[angle=270,width=3in]{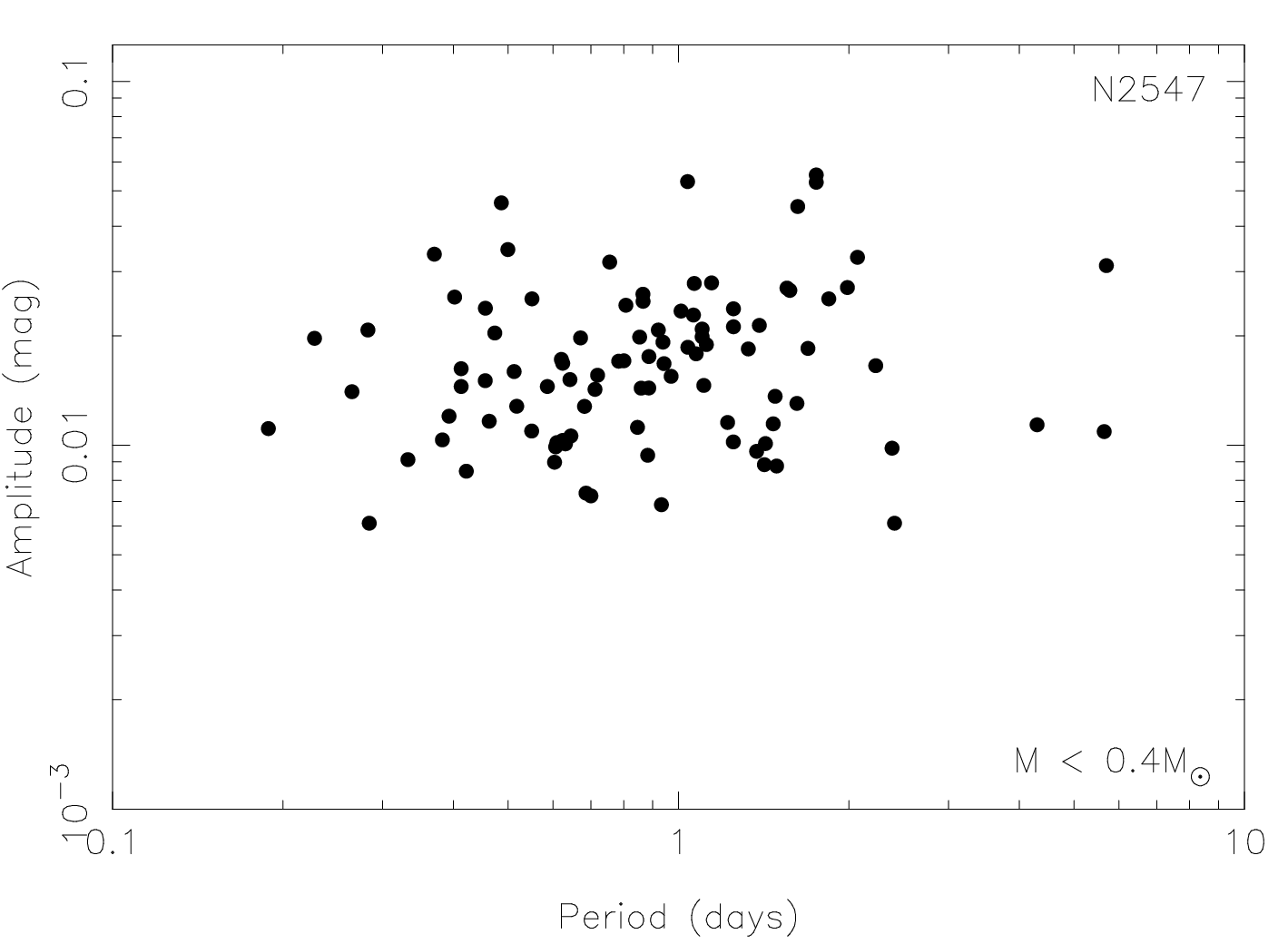}

\caption{Plots of amplitude as a function of period for NGC 2547,
  in two mass bins: $0.4 \le M/\msun < 1.0$ (top) and $M < 0.4\ \msun$
  (bottom).}

\label{pad}
\end{figure}

Could the apparent morphology in Figure \ref{pcd} be explained by
sample biases?  The simulations of \S \ref{sim_section} suggest that
this is unlikely, since we are sensitive to shorter periods than the
$0.2 - 0.3\ {\rm day}$ `limit', and the slight bias toward detection
of shorter periods at low mass is not sufficient to explain the
observations for the slow rotators.  Furthermore, Figure \ref{pad}
indicates that the lack of sensitivity to low amplitudes at low masses
does not appear to introduce any systematic changes in the detected
periods.

Figure \ref{pad} shows an apparent lack of objects with masses $M <
0.4\ \msun$ and rotation periods $\ga 2\ {\rm days}$, a region of
the diagram where the survey is sensitive down to amplitudes of $0.01\
{\rm mag}$ (see also \S \ref{sim_section}).

A very large fraction ($\sim 50\%$) of the bright rotators ($M \ga
0.4\ \msun$) were detected in the X-ray survey of \citet{j2006}.
Detection in X-rays is suggestive of youth and hence cluster
membership, which confirms that the level of field contamination is
low in the rotation sample, especially when the incomplete spatial
overlap between our optical survey and the XMM-Newton observation of
\citet{j2006} is accounted for.  Very few of the lower-mass stars in
Figure \ref{pcd} were detected in X-rays (e.g. for $M < 0.3\ \msun$
there was only a single X-ray detection), but this is not surprising
since these objects are fainter.  Indeed, Figure 3 of \citet{j2006}
indicates that the X-ray detection limit for a cluster member
corresponds to $I \sim 16$, equivalent to $M \sim 0.34\ \msun$ with
the models we are using.

\subsubsection{Period distributions}
\label{perioddist_section}

\begin{figure*}
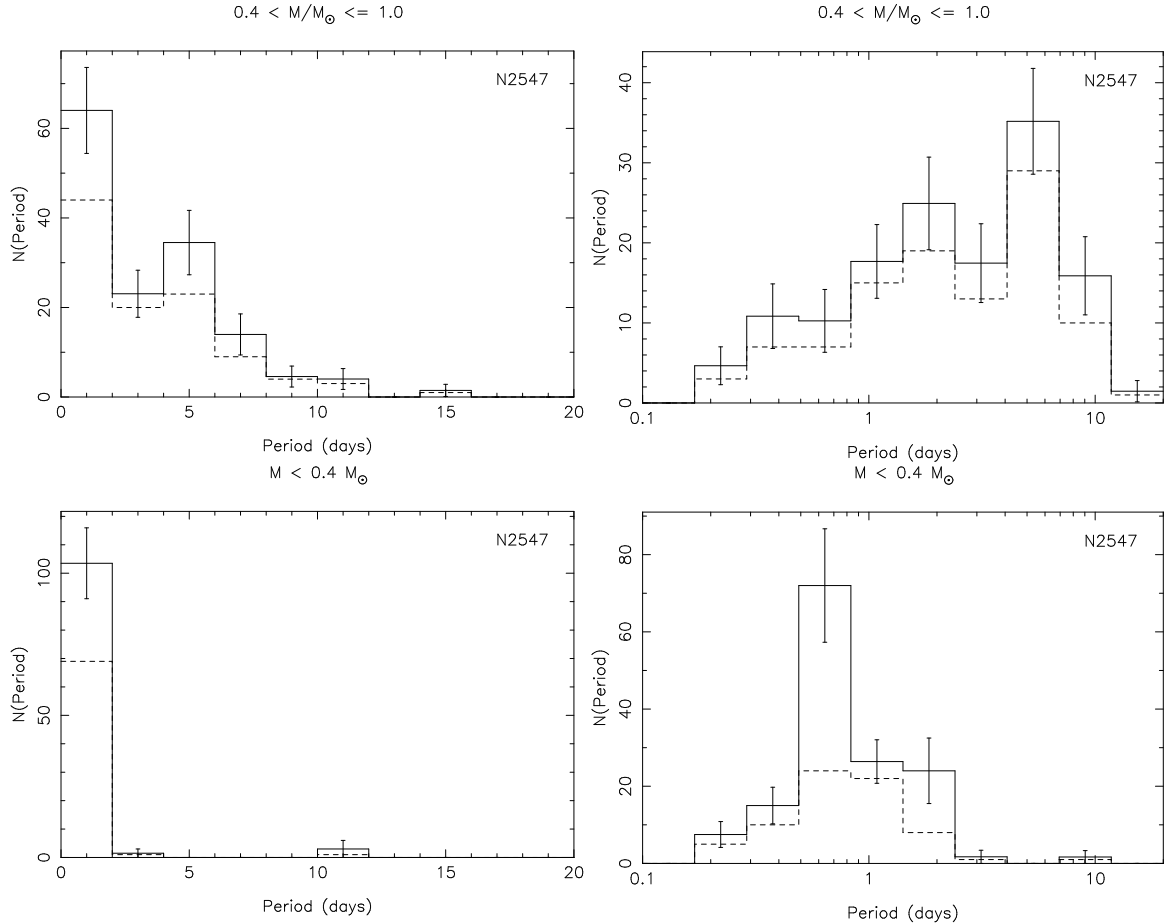

\centering
\includegraphics[angle=270,width=3in]{perioddist_1.ps}
\includegraphics[angle=270,width=3in]{perioddist_3.ps}
\includegraphics[angle=270,width=3in]{perioddist_2.ps}
\includegraphics[angle=270,width=3in]{perioddist_4.ps}

\caption{Period distributions for objects classified as possible
  photometric members, in two mass bins: $0.4 \le M/\msun < 1.0$
  (upper row, corresponding roughly to K and early-M spectral types)
  and $M < 0.4\ \msun$ (lower row, late-M).  The left-hand panels
  show the distributions plotted in linear period, and the right-hand
  panels show the same distributions plotted in $\log_{10}$ period.
  The dashed lines show the measured period distributions, and the
  solid lines show the results of attempting to correct for
  incompleteness and reliability, as described in the text.}

\label{perioddist}
\end{figure*}

In order to quantify the morphology of Figure \ref{pcd}, we have used
histograms of the rotation period distributions in two broad mass bins,
$0.4 \le M/\msun < 1.0$ and $M < 0.4\ \msun$, shown in Figure
\ref{perioddist}.  We have attempted to correct the distributions for
the effects of incompleteness and (un)reliability using the
simulations described in \S \ref{sim_section}, following the method
used in \citet{i2006}.  The results of doing this are shown in the
solid histograms in Figure \ref{perioddist}, and the raw period
distributions in the dashed histograms.

The period distributions in the two mass bins of Figure
\ref{perioddist} show clear differences, with the low-mass stars ($M <
0.4\ \msun$) showing a strongly peaked rotational period distribution,
with a maximum at $\sim 0.6-0.7\ {\rm days}$, whereas the higher-mass
stars ($0.4 \le M/\msun < 1.0$) show a broader distribution.  We
applied a two-sided Kolmogorov-Smirnov test to the corrected
distributions to confirm the statistical significance of this result,
finding a probability of $2 \times 10^{-13}$ that the distributions
were drawn from the same parent population.

The implication of this result is that the observed morphology in
Figure \ref{pcd}, and in particular the increase of the longest
observed rotation period as a function of increasing mass, a trend
followed also by the bulk of the rotators, is real and statistically
significant.

\subsubsection{Rapid rotators}

We have examined the periods of our fastest-rotating stars, to check
if they are rotating close to their break-up velocity.  The critical
period $P_{\rm crit}$ for break-up is given approximately by:
\begin{equation}
P_{\rm crit} = 0.116\ {\rm days}\ {(R / {\rm R_{\odot}})^{3/2}\over{(M / {\rm M_{\odot}})^{1/2}}}
\end{equation}
where $R$ and $M$ are the stellar radius and mass respectively
(e.g. \citealt{h2002}).  Using the NextGen models of \citet{bcah98},
the object rotating closest to breakup is N2547-2-2-1563, with $P_{\rm
  crit} / P = 0.46$.  However, the majority of our objects are
rotating at much lower fractions of their break-up velocity.

Figure \ref{pad} indicates that there is only one object with $P <
0.2\ {\rm days}$, N2547-2-2-1485 at $P = 0.188\ {\rm days}$.
Therefore, in common with the other clusters we have studied, we find
a short-period limit of $\sim 0.15 - 0.2\ {\rm days}$, with no objects
found rotating faster than this rate.

\subsection{The rotation-activity relation}
\label{xray_section}

Observations of open clusters in the age range $50\ {\rm Myr}$ to
several Gyr have established an age-rotation-activity relationship,
where younger stars tend to be more rapidly rotating, and emit strong
X-rays (up to a saturation level), whereas older stars have spun-down
to lower rotation rates, and emit a smaller fraction of their flux in
X-rays (\citealt{j99}; \citealt{r2000}).  \citet{j2006} established
that G and K stars in NGC 2547 follow the same relationship between
X-ray activity and Rossby number established for field stars and old
clusters, but at saturated or super-saturated X-ray activity levels,
with $L_x / L_{\rm bol}$ values similar to those for T-Tauri stars in
the ONC, but an order of magnitude higher than in the Pleiades.  In
this section, we re-evaluate the rotation-activity relation using the
present NGC 2547 rotation period sample, which probes to lower masses
than the sample of \citet{j2006}.

\begin{figure}
\centering
\includegraphics[angle=270,width=3in]{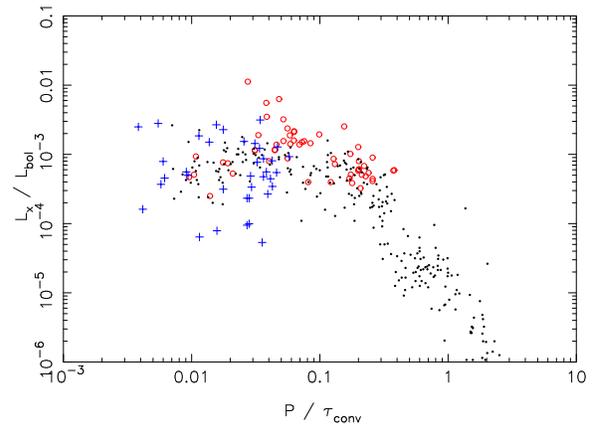}

\caption{Plot of X-ray activity ($0.1 - 2.4\ {\rm keV}$ luminosity
  divided by bolometric luminosity, as \citealt{j2006}) as a function
  of the Rossby number for the NGC 2547 rotation period data (open
  circles).  Also shown are the compilation of results for field and
  cluster stars (small black points) from \citet{piz2003}, and for
  ONC stars in the range $0.5 < M/\msun < 1.2$ (crosses) from
  \citet{getman2005}, assuming a fixed value of $\tau_c = 250\ {\rm
  yr}$ (see \citealt{preibisch2005}; \citealt{j2006}).}

\label{act}
\end{figure}

Following the method of \citet{j2006}, Figure \ref{act} shows a plot
of X-ray activity as a function of the Rossby number, defined as the
ratio of rotation period to convective overturn timescale at the base
of the convection zone, $\tau_{\rm conv}$.  Our X-ray luminosities
were taken from \citet{j2006}, but assuming a distance of $361\ {\rm
  pc}$ for consistency with the distance modulus of $7.79$ we assume
in this work (see \S \ref{intro_section}).  Bolometric luminosities
were calculated from the $V$-band absolute magnitudes and $V - I$
colours using a quadratic fit to the empirical bolometric corrections
and colours of \citet{l96}:
\begin{equation}
{\rm BC}_V = 0.243 - 0.583\ (V - I) - 0.128\ (V - I)^2
\end{equation}

The values of $\tau_{\rm conv}$ were derived using Eq. (4) of
\citet{n84} and $B - V$ colours of our stars, obtained from a linear
fit to the intrinsic stellar colours of \citet{leg92}:
\begin{equation}
(B - V) = 0.907 + 0.268\ (V - I)
\end{equation}

These relations should give reliable results over the entire
late-K to mid-M range of spectral types covered by the present
rotation sample.

Comparing Figure \ref{act} to Figure 9 of \citet{j2006} shows that we
reproduce the same trend of X-ray activity with Rossby number seen in
the earlier NGC 2547 work, and the ONC, where we have used a similar
plot scale to aid comparison with their Figure.

A small group of objects at $P / \tau_{\rm conv} \la 0.025$ and $L_X /
L_{\rm bol} \sim 10^{-3}$ in the NGC 2547 sample have unusually small
values of $P / \tau_{\rm conv}$ compared to the remainder of the NGC
2547 objects.  These all have very short rotation periods ($\la 0.6\
{\rm days}$).  Comparing with Figure 9 of \citet{j2006} indicates that
they also see one object in this region of the diagram.  It is
possible that these objects could show fast rotation rates as a result
of tidal synchronisation effects in binary systems, a hypothesis which
could be verified with follow-up spectroscopy.

There are also a small number of objects showing unusually strong
X-ray emission, with $L_X / L_{\rm bol}$ values up to $\sim 10^{-2}$,
which were not seen by \citet{j2006}.  Such high X-ray luminosities
would be indicative of very strong activity, or alternatively,
overestimation of $L_X$ or underestimation of $L_{\rm bol}$.  These
objects could also result from mis-identification of a very luminous
background AGN in X-rays with a cluster star in the optical, given the
large search radii (here $6''$) which must be used due to the limited
astrometric accuracy of the X-ray data. 

In order to attempt to trace the rotation-activity relation to
lower-mass in NGC 2547, it would be desirable to obtain deeper X-ray
observations, since the stars with masses $\la 0.35\ \msun$ were not
detected in the existing XMM-Newton observations of \citet{j2006}.

\subsection{Comparison with mid-IR \spitzer observations}

We have compared the rotation period results to the mid-IR \spitzer
observations of \citet{gorlova07}, to search for any correlation between
mid-IR excess (presumably due to a circumstellar disc) and rotation
period, as would be expected in the disc regulation paradigm for
angular momentum evolution on the early-PMS.

\begin{figure}
\centering
\includegraphics[angle=270,width=3in]{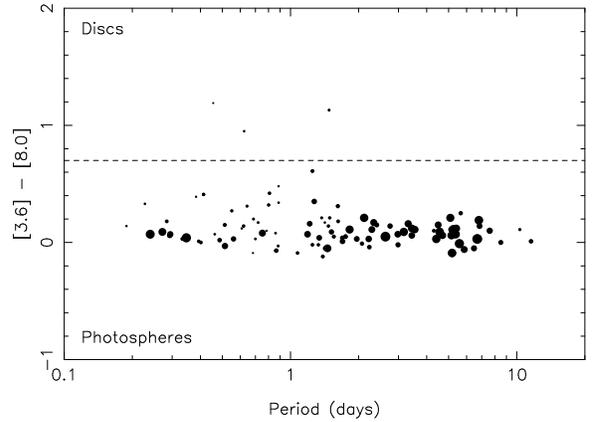}

\caption{Plot of $[3.6] - [8.0]$ colour as a function of period.  The
  sizes of the symbols indicate the masses of the objects, with the
  smallest symbols corresponding to $M \sim 0.1\ \msun$, and the
  largest to $M \sim 0.9\ \msun$.  The horizontal dashed line
  indicates indicates the threshold of $[3.6] - [8.0] > 0.7$ used by
  \citet{cb06} to select objects showing $8.0\ \micron$ excesses.}
\label{irex}
\end{figure}

Figure \ref{irex} shows a plot of $[3.6] - [8.0]$ colour versus
rotation period.  There are only three objects in the rotation period
sample identified as having possible $8.0\ \micron$ excesses.  These
are N2547-1-6-3669, N2547-2-2-6311, and N2547-2-3-6494.  Of these, the
latter two objects are flagged by \citet{gorlova07} as having possibly
unreliable $8.0\ \micron$ measurements, due to the presence of nearby
sources or other image features which might contribute to the measured
flux (confirmed by visual examination of our $I$-band images).
Therefore, only one object, N2547-1-6-3669, has a reasonably reliable
$8.0\ \micron$ excess.  This has a very low inferred mass of $0.21\
\msun$, and falls on the locus of slow rotators at this mass, having a
rotation period of $1.48\ {\rm days}$.

Given the extremely small number of objects under discussion, it is
difficult to draw any conclusions regarding the disc regulation
paradigm.  However, the only object with reliable evidence of an inner
disc, N2547-1-6-3669, is indeed a slow rotator, as would be expected
if its angular velocity was regulated by the presence of the disc.

One object in the rotation sample has a $24\ \micron$ detection,
and shows an excess in this band: N2547-1-5-1123.  Again, this has a
low inferred mass of $0.25\ \msun$, but is a rapid rotator, having a
period of $0.41\ {\rm days}$.  This object does not have an $8\
\micron$ excess, indicating that the inner disc has cleared, which
may also be consistent with the disc regulation paradigm, since the
inner disc is most likely the part that couples to the star to induce
angular momentum loss and prevent it from spinning up.

The existence of a relatively large fraction of $24\ \micron$
detections in the sample of \citet{gorlova07} motivates deeper MIPS
observations, to probe to lower masses: the existing observations were
only sufficient to detect photospheres down to F spectral types.  The
single object with a $24\ {\rm \micron}$ detection has a mid-M
spectral type, and was only detected as a result of its large $24\
{\rm \micron}$ excess.

\subsection{Comparison with other data-sets}

\subsubsection{Period versus mass diagram}
\label{pmd_section}

\begin{figure}
\centering
\includegraphics[angle=270,width=3in]{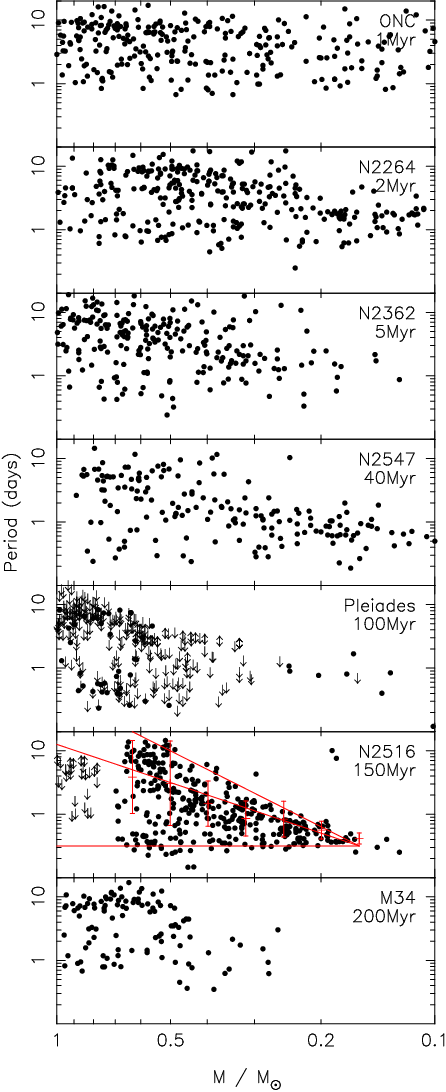}

\caption{Rotation period as a function of stellar mass for (top to
  bottom): ONC, NGC 2264, NGC 2362, NGC 2547, the Pleiades, NGC 2516
  and M34.  Lower and upper limits (from $v \sin i$ data) are marked
  with arrows. The masses were taken from the NextGen mass-magnitude
  relations \citep{bcah98} at the appropriate ages.  The ONC data are from
  \citet{h2002}.  For NGC 2264 we used the data of \citet{lamm05} and
  \citet{m2004}.  The NGC 2362 data are from the Monitor project, to
  be published in Irwin et al. (in preparation).  The Pleiades
  rotation period data are a compilation of the results from
  \citet{ple1}, \citet{ple2}, \citet{ple3}, \citet{ple4}, \citet{ple5},
  \citet{ple6}, \citet{ple7} (taken from the open cluster database),
  \citet{t99} and \citet{se2004b}.  The Pleiades $v \sin i$ data are a
  compilation of results from \citet{plev1}, \citet{plev2},
  \citet{plev3}, \citet{plev4}, \citet{plev5} and \citet{plev6}. The 
  NGC 2516 data are taken from \citet{i2007b}, plus upper limits from
  \citet{ter2002}, and the M34 data from \citet{i2006}.  In the NGC
  2516 plot, the lines show $P = {\rm constant}$, $P \propto M^2$ and
  $P \propto M^3$, and the error bars show the median of the
  distribution binned in $0.1\ {\rm dex}$ bins of $\log M$.}

\label{pmd}
\end{figure}

Figure \ref{pmd} shows a diagram of rotation period as a
function of stellar mass for the ONC ($1 \pm 1\ {\rm Myr}$;
\citealt{h97}), NGC 2264 ($2-4\ {\rm Myr}$; \citealt{park2000}), NGC
2362 ($\sim 5 \pm 1\ {\rm Myr}$; \citealt{moi2001}; \citealt{bl96}),
NGC 2547, the Pleiades ($\sim 100\ {\rm Myr}$; \citealt{mmm93}), NGC
2516 ($\sim 150\ {\rm Myr}$; \citealt*{jth2001}), and M34 ($\sim 200\
{\rm Myr}$; \citealt{jp96}).  Data sources for each cluster are
indicated in the figure caption.

The diagram clearly shows a gradual evolutionary sequence, from a
relatively flat mass-dependence of the rotation periods in the ONC
($\sim 1\ {\rm Myr}$), to a sloping relation in NGC 2362 ($\sim 5\
{\rm Myr}$) and NGC 2547, and the emergence of the break between a
flat distribution for $M \ga 0.6\ \msun$ and strongly sloping
distribution at lower masses, for the Pleiades, NGC 2516 ($\sim 150\
{\rm Myr}$) and M34 ($\sim 200\ {\rm Myr}$), as discussed in
\citet{i2007b}.

\section{Discussion}
\label{disc_section}

\subsection{Link to angular momentum}

\begin{figure}
\centering
\includegraphics[angle=270,width=3in]{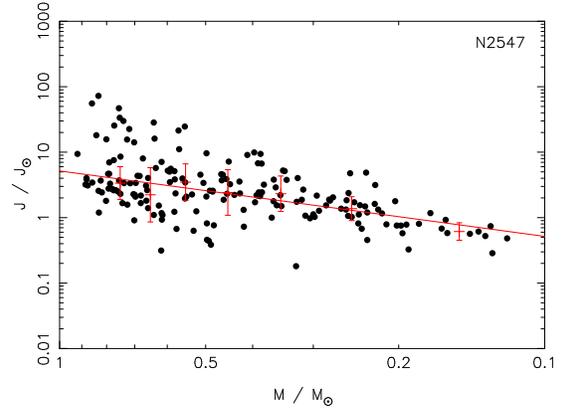}

\caption{Stellar angular momentum $J$ as a function of mass for the
  NGC 2547 rotation period sample.  The error bars show the median of
  the distribution binned in $0.1\ {\rm dex}$ bins of $\log M$, and
  the solid line shows a fit of $J \propto M$ to the median.}

\label{jmd_n2547}
\end{figure}

It is instructive to re-examine Figure \ref{pcd} in terms of the
stellar angular momentum $J$.  We define
\begin{eqnarray}
J &=& I \omega \\
  &=& {2 \pi k^2 M R^2 \over{P}}
\label{j_eqn}
\end{eqnarray}
where $I = k^2 M R^2$ is the moment of inertia of a star of
mass $M$ and radius $R$, $k$ is the radius of gyration ($k^2 = 2/5$
for a uniform sphere rotating as a solid body), and $\omega = 2 \pi /
P$ is the rotational angular velocity.  A plot of $J$ as a function of
stellar mass for the NGC 2547 sample is shown in Figure \ref{jmd_n2547}.

We computed the median angular momentum as a function of mass for the
NGC 2547 data, shown in Figure \ref{jmd_n2547}.  The relation is
consistent with the same $J \propto M$, or specific angular momentum
$j = {\rm constant}$, relation as seen by \citet{i2007b}, and
consistent with the conclusions of \citet*{h2001} in the ONC.  The
lack of any change in the median relation for stars still undergoing
contraction on the PMS indicates that any mass-dependent angular
momentum losses are not important in determining the evolution of the
median over this age range, unless high angular momentum is hidden in
a fast rotating radiative core.  However, this is not the case in
detail or once the stars reach the ZAMS, in particular it is clear
(e.g. from Figure \ref{pmd}) the {\em shape} of the distribution does
evolve in a mass-dependent fashion (see also the next Section).

\subsection{Simple models}
\label{model_section}

\subsubsection{Description}

For a full description of our simple model scheme, the reader is
referred to the NGC 2516 publication \citep{i2007b}.  Briefly, we have
generated models for solid body rotation, and a simple implementation
of core-envelope decoupling where the core and envelope are treated as
two separate entities joined at the convective/radiative boundary by
angular momentum transfer according to the prescription of
\citet{mac91}.  The angular momentum loss rate was split into two
components: losses due to stellar winds, assuming a loss law with
saturation at a critical angular velocity $\omega_{\rm sat}$ (allowed
to vary as a function of mass), and losses due to disc locking, which
to a good approximation maintains a constant angular velocity until
the circumstellar disc dissipates at an age $\tau_{\rm disc}$.  For
the decoupled models there is one further parameter, $\tau_c$, the
timescale for coupling of angular momentum between the core and
envelope.

For NGC 2547, we have computed the object masses using two different
values of the cluster parameters (particularly, the age), in order to
examine the consistency of the rotation periods with the range of
cluster ages found in the literature.  The first set of parameters are
those assumed earlier, corresponding to an age of $38.5\ {\rm Myr}$,
which is consistent with the Lithium age for the cluster of $\sim 35\
{\rm Myr}$ \citep{j2005}.  The second set of parameters corresponded
to the ``conventional'' isochrone based estimates, from \citet{n2002},
with an age of $25\ {\rm Myr}$, and corresponding distance modulus $(M
- m)_0 = 8.05$.

\subsubsection{Evolution from NGC 2362}
\label{ove_section}

We first attempt to evolve the observed rotation rates from our survey
in NGC 2362 (Irwin et al., in preparation) forward in time to
reproduce the available rotation period data.  Following the method
used in Section 6.2.2 of \citet{i2007b}, we have characterised the
slow rotator population by the $25$th percentile (the lower quartile)
of the distribution of observed angular velocities, $\omega$, and the
fastest rotators by the $90$th percentile.

\begin{figure*}
\centering
\includegraphics[angle=0,width=6in]{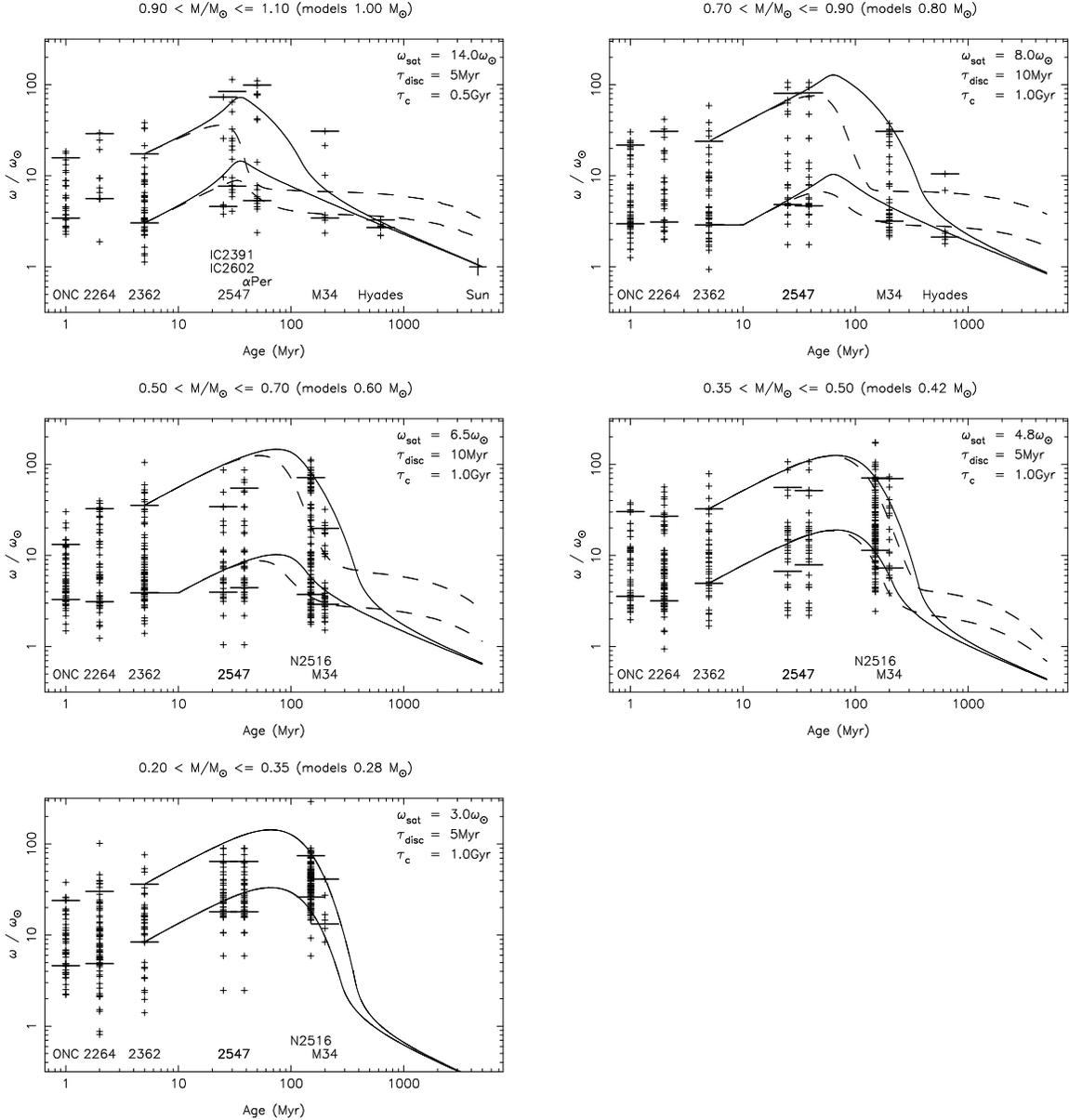}

\caption{Rotational angular velocity $\omega$ plotted as a function of
  time in five mass bins: $0.9 < M/\msun \le 1.1$, $0.7 < M/\msun \le
  0.9$, $0.5 < M/\msun \le 0.7$, $0.35 < M/\msun \le 0.5$, and $0.2 <
  M/\msun \le 0.35$.  Crosses show the rotation period data, and short
  horizontal lines the $25$th and $90$th percentiles of $\omega$, used
  to characterise the slow and fast rotators respectively.
  The lines show our models for $1.0$, $0.8$, $0.6$, $0.42$ and $0.28\
  \msun$ (respectively), where the solid lines are solid body models,
  and dashed lines are differentially rotating models, with the
  parameters shown.  Plotted are the ONC ($1\ {\rm Myr}$), NGC 2264
  ($2\ {\rm Myr}$), NGC 2362 ($5\ {\rm Myr}$), IC 2391, IC 2602 ($\sim
  30\ {\rm Myr}$), $\alpha$ Per ($\sim 50\ {\rm Myr}$), NGC 2547
  (plotted for ages of $25$ and $38.5\ {\rm Myr}$, as described in the
  text), M34, the Hyades ($625\ {\rm Myr}$) and the Sun ($\sim 4.57\
  {\rm Gyr}$).
  The IC 2391 data were taken from \citet{ps96} and IC 2602 from
  \citet{bsps99}.  The $\alpha$ Per data are a compilation of the
  results  from \citet{aper1}, \citet{aper2}, \citet{ple4},
  \citet{aper3}, \citet{ple5}, \citet{aper4}, \citet{ple6},
  \citet{aper5}, \citet{aper6}, \citet{aper7}, \citet{aper8},
  \citet{aper9}, \citet{aper10}, \citet{aper11}, and the Hyades data
  from \citet{hya1} and \citet{ple6}, taken from the open cluster
  database.}
\label{pevol}
\end{figure*}

Figure \ref{pevol} shows a revised version of Figure 19 of
\citet{i2007b}, adding in the NGC 2547 data.  We have re-used the
solid body and differentially rotating model fits to the NGC 2516
data: the models were {\it not} re-fit with the NGC 2547 observations.
The mass bins used were as before: $0.9 < M/\msun \le 1.1$, $0.7 <
M/\msun \le 0.9$, $0.5 < M/\msun \le 0.7$, $0.35 < M/\msun \le 0.5$,
and $0.2 < M/\msun \le 0.35$ (chosen empirically to encompass the
changes in behaviour seen in the models, while retaining reasonable
statistics).  The models were calculated for single masses roughly at
the centre of these bins, of $1.0$, $0.8$, $0.6$, $0.42$ and $0.28\
\msun$.  Since there are very few observations (one) in the
highest-mass bin for NGC 2547, this will not be discussed further
here.

Model parameters are summarised in each panel of Figure \ref{pevol}.
Briefly, we fit the models using NGC 2362 as an initial condition, to
best-reproduce the NGC 2516 rotation periods.  The saturation angular
velocity $\omega_{\rm sat}$ and disc lifetime $\tau_{\rm disc}$ were
varied to obtain the best-fit for the solid body models to the fast
rotators, and for the decoupled models we varied $\tau_c$, the
core-envelope coupling timescale, to fit the slow rotators.  See
\citet{i2007b} for a more detailed discussion.

The models for the $0.7 < M/\msun \le 0.9$ bin show overall good
agreement with the NGC 2547 data.  At this mass and age it is
difficult to distinguish between the solid body and decoupled models,
but at later ages, the solid body model appears to give a better fit
to the rapid rotators, and a differentially-rotating model to the
slow rotators \citep{i2007b}.

However, in the remaining three bins (and to some extent in the $0.7 < 
M/\msun \le 0.9$ bin), a clear trend is visible for both the fast and
slow rotator populations, where the observations show that the stars
in NGC 2547 are rotating slower than expected.

Returning to Figure \ref{pmd}, it is clear that the fastest rotators
have spun up significantly from NGC 2362 to NGC 2547, in accordance
with the model prediction.  Furthermore, there is some degree of
scatter in the NGC 2362 data, especially at short periods, which may
be due to contamination, e.g. from field binaries (Irwin et al., in
preparation) in a similar fashion as we have hypothesised in M34
\citep{i2006} and NGC 2516 \citep{i2007b}.  If this is true, the
presence of these objects has shifted the $90$th percentile in NGC
2362 upward in $\omega$, and may partly explain the discrepancy,
especially due to the small number of objects in question.  It is
difficult to confirm this hypothesis before obtaining follow-up
spectroscopy for the NGC 2362 sample to determine the nature of these
objects.

For the slow rotators, examining Figure \ref{pmd} indicates that there
are a number of `outliers' at longer period than the general trend
shown by the upper envelope of the rotators.  It is possible that
that if these objects were removed from the computation of the
percentiles, the slow rotators would then fall into agreement with the
models.  The nature of these objects is not clear, and we defer
discussion of them in detail until we can obtain spectroscopy to
confirm their nature as slow rotators, and confirm the masses we have
determined from the models of \citet{bcah98}.

Returning to the issue of the age of NGC 2547, the rotation periods
seem to be slightly more consistent with the younger $25\ {\rm Myr}$
age, although the results are still clearly not reproduced by the
models, especially in the intermediate-mass $0.5 < M/\msun \le 0.7$
and $0.35 < M/\msun \le 0.5$ bins, where the NGC 2547 objects appear
to be rotating slower than the models predict.  Moving the NGC 2547
points to the left in the diagram would improve agreement with the
models, but this is difficult to justify since the youngest ages in
the literature for this cluster are $\sim 20-25\ {\rm Myr}$.  Given
the uncertainties in the models, the rotation period measurements
cannot therefore place useful constraints on the cluster age.

In light of the discussion above, the NGC 2547 data seem to be in
reasonable agreement with the models, pending the results of
spectroscopic follow-up, without need for significant modification
from the earlier NGC 2516 study \citet{i2007b}.  At this age it is
difficult to observe the effects of core-envelope decoupling, so we
are unable to refute our earlier statements that core-envelope
decoupled models appear to be required to reproduce the evolution of
the slowest rotators, and that the same models do not appear to give a
good fit to the fast rotators, which are well-fit by solid body models.

\subsubsection{Detailed evolution}

\begin{figure*}
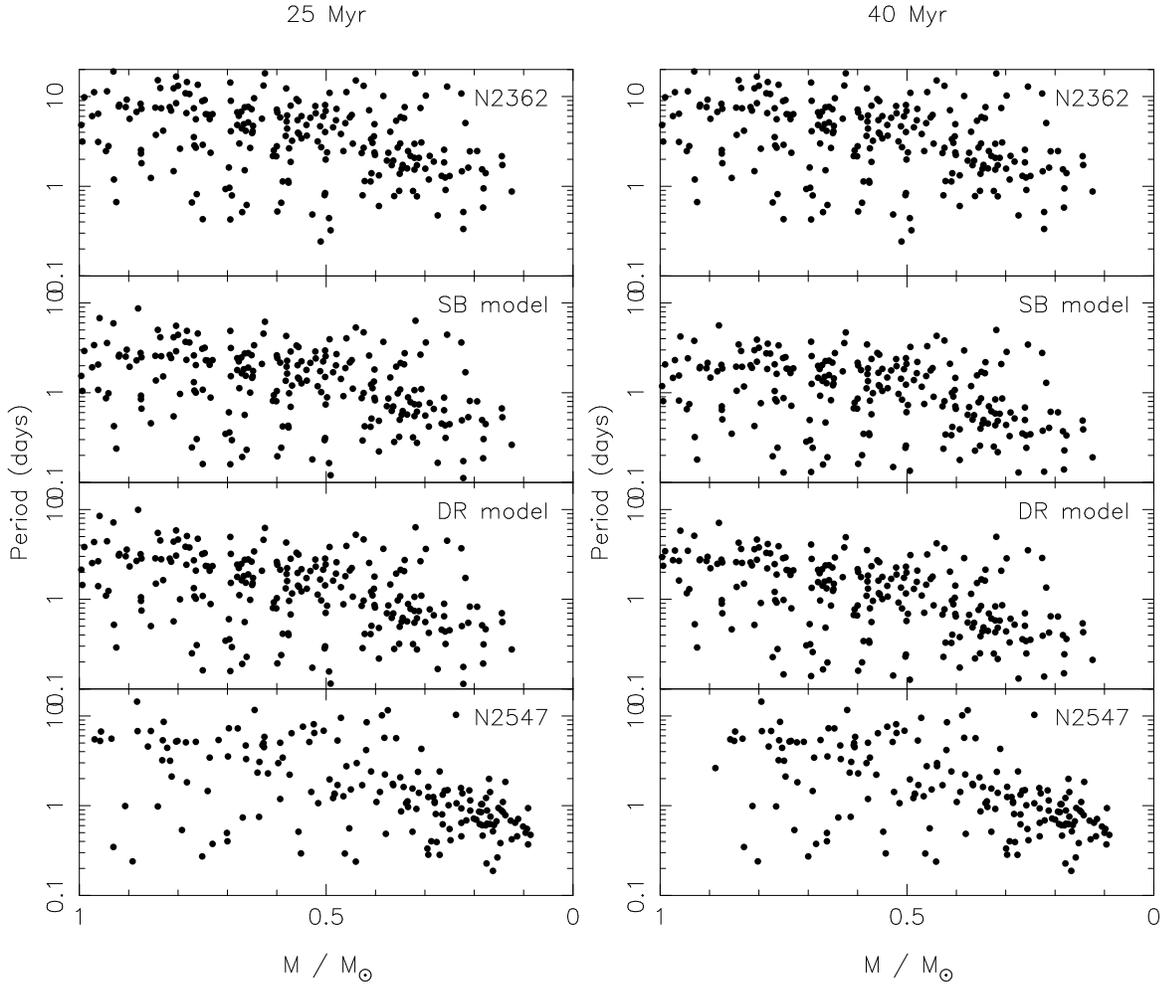

\centering
\includegraphics[angle=270,width=3in]{pmd_evol_25.ps}
\includegraphics[angle=270,width=3in]{pmd_evol_40.ps}

\caption{Rotation period as a function of mass, using the model
  presented in \S \ref{model_section} to evolve the NGC 2362
  distribution (top panel) forward in time from $5\ {\rm Myr}$ to $25\
  {\rm Myr}$ (left panels) and $40\ {\rm Myr}$ (right panels) for the
  solid body (second panel) and differentially rotating (third panel)
  models, and the observed NGC 2547 distribution for comparison
  (bottom panel).}

\label{jmd_n2362_evol}
\end{figure*}

Figure \ref{jmd_n2362_evol} shows the results of attempting to evolve
the measured rotation periods in NGC 2362 forward in time to the age
of NGC 2547 using the models we have described, with the relations fit
from the NGC 2516 data for $\omega_{\rm sat}$ as a function
of mass, on an object-by-object basis.  By doing this, we can test if
the model we have presented can reproduce the observations of NGC
2547, given the NGC 2362 rotators as an input.

Comparing the left and right-hand panels in Figure
\ref{jmd_n2362_evol} indicates that the results are relatively
independent of the assumed age for NGC 2547.  This is due to the
relatively shallow slope of the time-dependence in the rotation rates
during this age range (see Figure \ref{pevol}).

The shape of the mass dependent morphology of the NGC 2547
distribution is reasonably well-reproduced from the NGC 2362
distribution by the models, with the differentially rotating models
providing a slightly better fit overall to the slow rotators, but at
this age it is difficult to distinguish between the two classes of
models since the effects of differential rotation become much more
pronounced between $\sim 50-100\ {\rm Myr}$.

Furthermore, the rotation rates are also well-reproduced below $\sim
0.7\ \msun$, but not for higher masses, where the predicted rates are
$\sim$ a factor of two faster than actually observed in NGC 2547.  It
should be noted that the adopted values of $\omega_{\rm sat}$ in this
mass range are not well-constrained by the NGC 2516 data used to
calibrate them, and have a strong effect on the morphology of this
part of the diagram, so it is possible that any deviations here result
from our assumption of a linear relation over the entire mass range.

The models predict that a number of objects should be rotating faster
than $\sim 0.2\ {\rm days}$, whereas none were observed in NGC 2547.
However, re-examining Figure \ref{pmd} shows some scatter in the NGC
2362 distribution at short periods, which may be related to
contamination of the sample, so it is difficult to conclude if this
result is significant before obtaining follow-up of the NGC 2362
sample.

This work suggests overall that the mass dependence of the observed
rotation period distribution is reasonably well-reproduced by the
combination of the initial conditions as seen in NGC 2362, the
mass-dependence of stellar contraction on the PMS, and the
mass-dependence of the saturation angular velocity $\omega_{\rm sat}$,
in agreement with \citet{i2007b}.

\section{Conclusions}
\label{conclusions_section}

We have reported on results of an $I$-band photometric survey of NGC
2547, covering $\sim 0.6\ {\rm sq. deg}$ of the cluster.  Selection
of candidate members in a $V$ versus $V-I$ colour-magnitude diagram
using an empirical fit to the cluster sequence found $800$
candidate members, over a $V$ magnitude range of $12.5 < V < 24$
(covering masses from $0.9\ \msun$ down to below the brown dwarf
limit).  The likely field contamination level was estimated using a
simulated catalogue of field objects from the Besan\c{c}on Galactic
models \citep{r2003}, finding that $\sim 470$ objects were likely
field contaminants, an overall contamination level of $\sim 59 \%$,
implying that there are $\sim 330$ real cluster members over this
mass range in our field-of-view.

From $\sim 100\ {\rm hours}$ of time-series photometry we derived
light curves for $\sim 130\,000$ objects in the NGC 2547 field,
achieving a precision of $< 1 \%$ per data point over $14 \la I \la 18$. 
The light curves of our candidate cluster members were searched for
periodic variability corresponding to stellar rotation, giving $176$
detections over the mass range $0.1 < M/\msun < 0.9$.

The rotation period distribution a a function of mass was found to
show a clear mass-dependent morphology, intermediate between those
found in NGC 2362 (Irwin et al., in preparation) and NGC 2516
\citep{i2007b}, with a median relation of $J \propto M$, or $j = {\rm
  constant}$, as seen in the other clusters we have surveyed.

In \S \ref{model_section}, simple models of the rotational evolution
were considered, both for the solid body case, and including
differential rotation between a decoupled radiative core and
convective envelope, as in our earlier work on NGC 2516
\citep{i2007b}.  Allowing for the uncertainty in the age of NGC 2547,
the results indicate that it is possible to qualitatively reproduce
the shape of the observed rotation rate distribution by evolving the
NGC 2362 distribution forward in time, using the parameters fit to the
NGC 2516 data of \citet{i2007b}.  It is difficult to examine the
effects of differential rotation at this age since they are still
relatively small. Some small discrepancies were noted, caused by what
appear to be `outliers' in the distributions for NGC 2362 and NGC
2547, but it is difficult to resolve these issues at the present time.
Follow-up spectroscopy will be required to make further progress.

\section*{Acknowledgments}

Based on observations collected at the European Southern Observatory,
Chile, as part of ESO large program 175.C-0685.  This publication
makes use of data products from the Two Micron All Sky Survey, which
is a joint project of the University of Massachusetts and the Infrared
Processing and Analysis Center/California Institute of Technology,
funded by the National Aeronautics and Space Administration and the
National Science Foundation.  This research has also made use of the
Digitized Sky Surveys, which were produced at the Space Telescope
Science Institute under U.S. Government grant NAG W-2166, the SIMBAD
database, operated at CDS, Strasbourg, France, and the WEBDA database,
operated at the Institute for Astronomy of the University of Vienna.
The Open Cluster Database, as provided by C.F. Prosser and
J.R. Stauffer, may currently be accessed at {\tt
  http://www.noao.edu/noao/staff/cprosser/}, or by anonymous ftp to
{\tt 140.252.1.11}, {\tt cd /pub/prosser/clusters/}.

JI gratefully acknowledges the support of a PPARC studentship, and SA
the support of a PPARC postdoctoral fellowship.  We would like to
express our gratitude to Isabelle Baraffe for providing the stellar
evolution model tracks used in \S \ref{model_section}, and to Erick
Young for supplying a machine-readable table of the \spitzer
mid-IR observations.

\end{document}